\documentstyle[11pt,appb,axodraw]{article}

    \advance\textheight by \topskip

\def\sign(#1){(\!-\!1)^{#1}}
\def\binom(#1,#2){ (\!\!
	 \begin{array}{c} #1 \\ #2 \end{array}\!\! ) }
\def\plus{\!+\!}
\def\minus{\!-\!}

\def\nn{\nonumber \\ &&}
\def\Tr{{\rm Tr~}}
\def\Str{{\rm Str~}}
\def\TB(#1,#2,#3,#4,#5){
	\raisebox{-19.1pt}{
	\SetScale{0.5} \SetPFont{Helvetica}{14}
	\hspace{-15pt}
	\begin{picture}(50,39)(0,-4)
	\SetColor{Blue}
	\CArc(40,35)(25,90,270) \CArc(60,35)(25,270,90)
	\Line(40,60)(60,60) \Line(40,10)(60,10) \Line(50,10)(50,60)
	\Line(0,35)(15,35) \Line(85,35)(100,35)
	\SetColor{Black}
	\PText(53,40)(0)[l]{#5}
	\PText(35,62)(0)[rb]{#1} \PText(65,62)(0)[lb]{#2}
	\PText(65,12)(0)[lt]{#3} \PText(35,12)(0)[rt]{#4}
	\end{picture}
	\SetScale{1.0}
	\hspace{-7pt}
	}
}
\def\TAH(#1,#2,#3,#4,#5,#6,#7){
	\raisebox{-19.1pt}{
	\SetScale{0.5} \SetPFont{Helvetica}{14}
	\hspace{-15pt}
	\begin{picture}(50,39)(0,-4)
	\SetColor{Blue}
	\CArc(40,35)(25,90,270) \CArc(60,35)(25,270,90)
	\Line(40,60)(60,60) \Line(40,10)(60,10) \Line(50,10)(50,60)
	\Line(0,35)(15,35) \Line(85,35)(100,35)
	\SetColor{Black}
	\PText(40,65)(0)[rb]{#1}
	\PText(60,65)(0)[lb]{#2}
	\PText(65,12)(0)[lt]{#3}
	\PText(35,12)(0)[rt]{#4}
	\PText(53,40)(0)[l]{#5}
	\PText(15,48)(0)[rb]{#6}
	\PText(85,48)(0)[lb]{#7}
	\SetColor{Red}
	\SetWidth{3}
		\Line(40,60)(60,60)
		\CArc(40,35)(25,90,130)
		\CArc(60,35)(25,50,90)
	\SetWidth{0.5}
	\end{picture}
	\SetScale{1.0}
	\hspace{-7pt}
	}
}
\def\TAZ{
	\raisebox{-19.1pt}{
	\SetScale{0.5} \SetPFont{Helvetica}{14}
	\hspace{-15pt}
	\begin{picture}(50,39)(0,-4)
	\SetColor{Blue}
	\CArc(40,35)(25,90,270) \CArc(60,35)(25,270,90)
	\Line(40,60)(60,60) \Line(40,10)(60,10) \Line(50,10)(50,60)
	\Line(0,35)(15,35) \Line(85,35)(100,35)
	\SetColor{Red}
	\SetWidth{3}
		\CArc(40,35)(25,130,230)
	\SetWidth{0.5}
	\end{picture}
	\SetScale{1.0}
	\hspace{-7pt}
	}
}

\begin{document}
\hfill
NIKHEF-98-021
\begin{center}
{\huge \bf Some problems in loop calculations } \\[8mm] 
J.A.M. Vermaseren \\ [3mm]
NIKHEF, P.O. Box 41882, \\ 1009 DB, Amsterdam, The Netherlands \\
\end{center}

\begin{abstract}
We discuss some of the problems that may occur in the calculation of 
complicated Feynman diagrams. These include the group independent 
evaluation of color factors, and the summation techniques that are needed 
for the expansion of diagrams into their Mellin moments.
\end{abstract}

\section{Introduction}
The ever increasing accuracy of the high energy experiments forces 
theorists to calculate perturbative quantities to higher and higher order. 
For high order graphs there are however not only the integrals to worry 
about. One of the problems is the organization of calculations with 
astronomical numbers of diagrams. This we will not discuss here. Another 
problem concerns the color factors. Traditionally they have been evaluated 
for QCD with an algorithm that was specific for the SU(N) 
groups~\cite{cvitanovic}~\cite{schladming}. These results could then be 
rewritten in terms of Casimir invariants and applied to other groups as 
well. This procedure works well as long as the only relevant invariants are 
$C_F$ and $C_A$. Recent calculations (e.g. ref.~\cite{QCD4}) however went 
beyond this and needed additional invariants. Hence new algorithms were 
needed. We will discuss them here.

Another problem that arises concerns the hadron structure functions in deep 
inelastic scattering. Here the structure functions can be computed to two 
loops but for a complete NNLO analysis one will also need the three loop 
anomalous dimensions. Of these anomalous dimensions only some Mellin 
moments are known thus far. This raises several points simultaneously. The 
first is the computation of the Mellin moments, once the exact result in 
x-space is known. The second problem is that one would like to use these 
moments for at least a partial analysis. And thirdly one may wonder whether 
it could be possible to evaluate all moments simultaneously. Or in other 
words: can one compute these diagrams in Mellin space? This technique has 
been used in the past for several 
calculations~\cite{First}~\cite{Madrid}~\cite{KandK}, but each time the 
calculations were done mainly by hand, in which case one can use many 
tricks to come to an answer. Additionally these techniques have never been 
used to the level of complexity one needs for the anomalous dimensions in 
NNLO. In the case of large numbers of diagrams one will have to solve the 
problem rather thoroughly. We will discuss some of this and show some 
progress in at least the first point.

\section{Color factors}
The evaluation of color factors is a purely mathematical problem. Yet most 
of the literature about it has been written by physicists. The most widely 
used paper is the one by Cvitanovic~\cite{cvitanovic}. In this paper 
explicit properties of the fundamental and adjoint representations of 
various groups are used to break the traces down and obtain expressions for 
the traces in terms of N (for SU(N), SO(N) and Sp(N)) or just numbers (for 
$G_2$, $F_4$, $E_6$ and $E_7$). The algorithms for the exceptional groups 
are difficult to implement and for $E_8$ no algorithms are given. But if 
one constructs a program for SU(N) it can be rather fast, and for `simple' 
QCD calculations this is all one needs, because the quarks are in the 
fundamental representation and the gluons are in the adjoint 
representation. Such programs can be quite short in any symbolic 
manipulation language (see for instance ref. ~\cite{schladming}). For most 
calculations thus far this is all one needs because for QED and SU(2)xU(1) 
the group theory is completely trivial. Nowadays however there is much 
activity concerning what lies beyond the standard model, and a much larger 
variety of groups can occur. Hence it would be wise to present perturbative 
calculations in such a way that the group and the representation(s) have 
not been fixed yet. Yet one would like to have a compact result. The 
essence of such an endeavor is or course to utilize no group or 
representation specific information, and to express the result in terms of 
as few invariants as possible. These invariants then can be either 
tabulated for different groups and representations, or there should be easy 
algorithms to evaluated them. What we are going to show here is a very 
short version of a recent paper~\cite{kleur} on this subject. For more 
details the reader should consult the original paper.

First we take
\begin{equation}
      [T^a_R, T^b_R] = i\ f^{abc}T^c_R
\end{equation}
Of special interest are two quadratic Casimir operators:
\begin{eqnarray}
	(T_R^a T_R^a)_{ij} & = & C_R\delta_{ij} \\
	f^{acd}f^{bcd} & = & C_A\delta^{ab}
\end{eqnarray}
The index $R$ in $T_R$ and $C_R$ labels the representation. The next set of 
invariants we consider are symmetrized traces.
\begin{equation}
   \Str T^{a_1}\ldots T^{a_n} \equiv {1 \over n!} \sum_{\pi} 
    \Tr T^{a_{\pi(1)}}\ldots  T^{a_{\pi(n)}}
\end{equation}
For each representation one may define a symmetric invariant tensor $d_R$ 
with
\begin{equation}
    d_R^{a_1\ldots a_n}\equiv \Str T_R^{a_1}\ldots  T_R^{a_n}
\end{equation}
This does overparametrize the problem, but if we manage to express all 
color factors in terms of contractions of such objects we have made an 
enormous simplification.

For a reduction into invariants we first deal with all generators that are 
not in the adjoint representation. If there are still open indices we 
multiply with appropriate projection operators. This gives complete traces 
and we have to find an expression for them in terms of symmetrized traces. 
Of course one can eliminate contracted indices inside the same trace as one 
does with traces over gamma matrices. Here this is more complicated though. 
Writing the expression in terms of symmetrized traces can either be done by 
recursion or with a closed formula which is a bit messy to write here. The 
formula gives much faster results, but in both cases the expressions become 
rather long when there are many generators in the trace.

Next one has an expression with invariants $d_R$ and `structure constants' 
$f$ which can be considered as proportional to generators of the adjoint 
representation. These do not necessarily occur in loops. If they do, they 
can be written also in terms of invariants $d_A$. After this step we are 
left with combinations of invariants and structure constants in which the 
structure constants cannot be arranged in terms of loops. At this point we 
start using Jacobi identities. For a computer this is not easy, but the 
program we constructed can do this for color traces that contain (at the 
beginning) up to 14 generators without any problem and for 16 it works 
almost always. This manages to reduce all such traces to contractions 
between invariants $d$ with the exception of one combination of three 
tensors $d_R$ and two structure constants $f$ when we started with 14 
generators, and three similar objects at the 16 generator level. In some 
cases these objects can be reduced (like if at least two of the three 
invariants $d$ belong to the adjoint representation, or for groups for 
which some identities hold which includes almost, but unfortunately not all 
groups). The resulting objects are considered as fundamental by the 
program. It is possible to express them in terms of an even smaller set of 
independent objects, but unfortunately this will not make the expressions 
shorter, because the constants in such a reduction are not simple. Hence 
the program gives its answers in these objects and one can evaluate these 
contractions afterwards for any given group. The paper presents also the 
formalism on how to do this for each group and for any given 
representation. Here we just give some examples of some rather complicated 
color traces. All examples use an experimental version of the program FORM 
which will be released later this year.

We first look at the following trace:
\[ R_{nn} = Tr [ T_R^{i_1}\cdots T_R^{i_n}T_R^{i_1}\cdots T_R^{i_n} ] \]
which gives some type of maximal complexity. For $n=7$ we obtain:
\begin{eqnarray}
	R_{77} & = &
	112/3\ d_R^{abcdef}d_A^{abcg}d_A^{defg}
	-328/9\ d_A^{abcdef}d_R^{abcg}d_A^{defg} \nn
  +d_R^{abcdef}d_A^{abcdef}(-56\ C_R+296/3\ C_A) \nn
 +d_R^{abcd}d_A^{abef}d_A^{cdef}(42\ C_R-749/10\ C_A)
    +67/15\ I_2(R)d_A^{abcd}d_A^{abef}d_A^{cdef} \nn
 +d_R^{abcd}d_A^{abcd} (35\ C_R^3-357/2 C_R^2C_A+868/3\ C_RC_A^2
		-2695/18\ C_A^3) \nn
 +I_2(R)d_A^{abcd}d_A^{abcd} (7\ C_R^2-1603/60 C_RC_A+497/20\ C_A^2 )
		 \nn
 +N_AI_2(R)(
       + C_R^6
       - 21/2        \ C_R^5C_A  
       + 175/4       \ C_R^4C_A^2
       - 280/3       \ C_R^3C_A^3 \nn
       + 5215/48     \ C_R^2C_A^4
       - 19075/288   \ C_R  C_A^5
       + 43357/2592  \ C_A^6 )
\end{eqnarray}
in which $N_A$ is the dimension of the adjoint representation and $I_2(R)$ 
is the second index of the representation $R$ and it can also be written as
$I_2(R) = (N_R C_R)/N_A$ with $N_R$ the dimension of the representation 
$R$. This computation took less than 35 sec on a PP200 running NeXTstep.

If the representation $R$ in the above example is taken to be the adjoint 
representation things are much simpler and much quicker:
\begin{eqnarray}
	A_{77} & = & 
		-\frac{8}{9}d_A^{abcdef}d_A^{abcg}d_A^{defg}
		+\frac{53}{30}C_A\ d_A^{abcd}d_A^{abef}d_A^{cdef}
		-\frac{5}{648}N_AC_A^7
\end{eqnarray}
and the computer time needed is less than 1 sec. Finally a topologically 
complicated example: It is called the Coxeter graph. It contains 14 
vertices and the smallest loop in it has 6 vertices. 
\begin{eqnarray}
	G_6(n=14) & = &
		\frac{16}{9}d_A^{abcdef}d_A^{abcg}d_A^{defg}
		-\frac{8}{15}C_A\ d_A^{abcd}d_A^{abef}d_A^{cdef}
		+\frac{1}{648}N_AC_A^7
\end{eqnarray}
This took 1.6 sec.

\section{Sums}

The method of evaluating structure functions in Mellin space dates back to 
the origins of QCD~\cite{First} It has been used~\cite{Madrid} to obtain 
the anomalous dimensions of the deep inelastic structure functions at the 
two loop level. Kazakov and Kotikov~\cite{KandK} used it for obtaining the 
ratio $R = \sigma_L/\sigma_T$ of deep inelastic structure functions at the 
two loop level. A complete program for all two loop calculations has 
however not been constructed thus far. Hence there is a challenge here. The 
main prize would of course be a program that can compute the three loop 
anomalous dimensions. We will address here a project in which first a 
program is constructed to evaluate all relevant two loop coefficient 
functions in Mellin space. Then one has to look at the inverse Mellin 
transform to obtain results in x-space. Next would be the study of three 
loop anomalous dimensions. Thus far we cannot say much of this.

Let us have a look at a typical two loop diagram.
\begin{eqnarray}
	\TAH({A},{B},{c},{d},{e},{a},{b}) & = & \int d^Dp_1\ d^Dp_2
		\frac{1}{(p_1^2)^a((P\plus p_1)^2)^A(p_2^2)^b((P\plus p_2)^2)^B
		(p_3^2)^c(p_4^2)^d(p_5^2)^e}
\end{eqnarray}
We can attack this diagram in a variety of ways. The method one might 
prefer is the `brute force' method. One decides to compute the $N$-th 
moment in which we keep $N$ symbolic. Therefore the two denominators are 
expanded, which results in a single symbolic sum. The resulting integral 
can be attacked with the standard techniques, but one has to introduce 4 
more sums. The only good news is that no individual term has more than 4 
nested sums. Amazingly enough these sums can be solved, although one has to 
do quite some work teaching the computer summation. The summation packages 
of the big computer algebra programs are almost useless here. One runs 
however immediately into trouble when trying to solve a similar toplogy 
\TAZ. Thus far the brute force method has failed on this diagram. Hence one 
has to be a bit smarter. If one looks in the Kazakov and Kotikov paper one 
can see a number of reduction schemes by which relations between the 
various topologies are derived. Two of them can then be marked as the 
simplest ones. These are then evaluated. They need only a single sum if 
enough preperatory work is done. Then a next level of topologies can be 
done which gives sums that involve these simpler topologies. But because 
the whole scheme is properly built up, the sums do not mix in such a way 
that we run into very complicated sums. Still one may need a number of sums 
that are not readily available and hence quite some attention has to go 
into the construction of a program that can handle all available sums. 
Because the complete program is still under construction I cannot show too 
many details here but yet, some may be interesting.

The class of functions that we run into is called 'harmonic series'. What 
is shown here about them can be found in a more complete version in 
ref~\cite{summer}. There 
exists a variety of notations for them. Because the more conventional 
notation is not very useful for computer programs we use a slight 
modification of this notation. The basic function is
\begin{eqnarray}
S_m(n) & = & \sum_{i=1}^n\frac{1}{i^m}\ \ \ \ \ \ \ \ \ \ \ \! m > 0 \\
       & = & \sum_{i=1}^n\frac{\sign(i)}{i^m}\ \ \ \ \ \ \ \ m < 0
\end{eqnarray}
and higher functions are defined by recursion:
\begin{eqnarray}
S_{m_1,\cdots ,m_k}(n) & = & \sum_{i=1}^n\frac{1}{i^{m_1}}
		S_{m_2,\cdots ,m_k}(i)\ \ \ \ \ \ \ \ \ \ \ \! m_1 > 0 \\
                     & = & \sum_{i=1}^n\frac{\sign(i)}{i^{m_1}}
		S_{m_2,\cdots ,m_k}(i)\ \ \ \ \ \ \ \ \ m_1 < 0
\end{eqnarray}
These functions appear, amoung others, when $\Gamma$-functions are expanded 
in terms of $\epsilon$. But they also pop up in sums of the type
\begin{eqnarray}
	\sum_{i=1}^n \sign(i) \binom(n,i) \frac{1}{n^3} & = &
			-S_{1,1,1}(n)
\end{eqnarray}
which are rather common in these calculations. Certain classes of sums can 
be evaluated to any complexity of the participating harmonic series. An 
example is combinations of $S_{\cdots}(n\minus i)$, $S_{\cdots}(i)$ and 
powers of $1/i$ as in
\begin{eqnarray}
\label{eq:nmii}
	\sum_{i=1}^{n\minus 1}\frac{1}{i^2}S_2(n\minus i)S_{-2,-1}(i) & = &
       - 3S_{2,-4,-1}(n) - 4S_{2,-3,-2}(n) + 4S_{2,-3,-1,1}(n)
       - 3S_{2,-2,-3}(n) \nn
       + 2S_{2,-2,-2,1}(n) + S_{2,-2,-1,2}(n) + 2S_{2,2,-2,-1}(n)
       + 2S_{2,3,-1,-1}(n) \nn
       - 4S_{3,-3,-1}(n) - 4S_{3,-2,-2}(n) + 4S_{3,-2,-1,1}(n) \nn
       + 4S_{3,1,-2,-1}(n) + 2S_{3,2,-1,-1}(n) - 3S_{4,-2,-1}(n)
\end{eqnarray}
and so on. There are also complicated sums that have thus far resisted 
generalization. This means that we cannot put in arbitrary harmonic series. 
An example is
\begin{eqnarray}
    \sum_{j=0}^n\binom(n,j)\binom(n\plus j,2\plus j)
                    \frac{\sign(j)}{(j\plus 2)^2}S_1(n\plus j) & = &
            \frac{1}{(n\plus 1)(n\plus 2)}(2S_1(n)S_1(n\minus 2)
                -S_2(n\minus 2) \nn -S_{-2}(n\plus 2)
				-S_{-2}(n\minus 2)
            +(\minus 1)^n(n^2\plus n\plus 3)\frac{(n\minus 2)!}{(n\plus 
			2)!} \nn
             -S_1(n\minus 2)(1\plus \frac{1}{n\minus 1}\plus 
				\frac{1}{n}
                -\frac{1}{n\plus 1}-\frac{1}{n\plus 2})\ )
\end{eqnarray}
The sums become also much more difficult when only a 
`partial range' is considered as in
\begin{eqnarray}
	\sum_{i=1}^m \sign(i) \binom(n,i) \frac{1}{n^3} & = & ???
\end{eqnarray}
One solution would be the introduction of a new type of functions with two 
variables. One would then construct the proper relations for these 
functions and hopefully, at the end of the complete calculation most or all 
would cancel. This approach may be necessary for the three loop anomalous 
dimension. At the two loop level it is not needed.

The sums that occur in the evaluation of diagrams in Mellin space are of 
course closely related to the sums that one runs into when making Mellin 
transforms of complete results in $x$-space. However, in the case of the 
transforms there is an extra class of sums: sums to infinity. These are in 
principle easier, but we need to derive some extra relations for them if we 
would like to be able to do them to a sufficient depth. Additionally one 
needs to know the values of the harmonic series at infinity. These values 
give us a number of constants that do not enter the problem if we calculate 
the diagrams directly in Mellin space. Hence their cancellation serves as a 
good check. For instance, the Mellin transform of $c_2$ in 
ref~\cite{WillyEllen} 
results in a formula with 154 terms. Because this is a new result it is 
shown in the appendix. 
It does take the program however only a few seconds to 
evaluate it. There is however an interesting spin-off. From the 
calculations in Mellin space, we know which classes of functions can occur 
in the Mellin transform. We can construct a basis in the space of these 
functions. If we have an equal number of functions in $x$-space of which 
the Mellin transforms span the space formed by this basis, we can do the 
inverse Mellin transform by just solving a linear set of equations. 
Additionally it tells us that there cannot be any relations between the 
functions in $x$-space that we used. Let us have a look at this basis. 
First we define the `level' of a term that involves an harmonic series. 
This level is the sum of the absolute value of its indices and to that we 
add the number of powers of denominators in the term. Hence the level of 
the argument of the sum in eq.~\ref{eq:nmii} is $2+2+3 = 7$. This is also 
the level of the terms on the right hand side of the equation. For two loop 
calculations we will need only functions of a level up to four. At level 
one there are only two functions: $S_1(n)$ and $S_{-1}(n)$. The function 
$1/n$ can be written as $S_1(n)-S_1(n\minus 1)$. The two terms in this 
expression correspond to the same function in $x$-space except for an 
overall factor $1/x$ for the second term. Hence we will not consider $1/n$ 
as an independent function. Additionally we do not have to worry about 
products of harmonic series with identical arguments. These can always be 
expressed in terms of sums of terms that have only single harmonic series 
as is shown in an example:
\begin{eqnarray}
S_{1,2}(n)\ S_{-1,1}(n) & = &
       - S_{-2,1,2}(n) - S_{-2,2,1}(n) + S_{-2,3}(n) + 2\ S_{-1,1,1,2}(n)
		\nn
       + S_{-1,1,2,1}(n) - S_{-1,1,3}(n) - S_{-1,2,2}(n) - S_{1,-3,1}(n)
		\nn
       + S_{1,-1,1,2}(n)+ S_{1,-1,2,1}(n) - S_{1,-1,3}(n) + S_{1,2,-1,1}(n)
\end{eqnarray}
At any given level $k$ greater than 1 there are three times as many 
functions than at the previous level. For each function at level $k\minus 
1$ one can construct a function at level $k$ by adding at the left an index 
1, an index -1, or by raising the leftmost index by one (in absolute 
value). Hence at level 4 there are 54 independent functions. Because we 
also need the lower functions one has to consider 80 functions in $x$-space 
before one can make a full inverse Mellin transform. An example is:
\begin{eqnarray}
S_{-1,1,2}(n\minus 1) & \rightarrow &
	\frac{{\rm Li}_3(1\minus x)}{1\plus x}
	-\frac{\ln(1\minus x)}{1\plus x}\zeta_2
	-\frac{1}{1\plus x}\zeta_3 \nn
	+\delta(1\minus x)(-\frac{1}{8}\zeta_4+\frac{1}{8}\zeta_3\ln(2)
		-\frac{1}{24}(\ln(2))^4 - {\rm Li}_4(\frac{1}{2})\ )
\end{eqnarray}

There is one case in which the basis of single higher harmonic series is 
not practical. This is when one has to evaluate these series at infinity. 
In that case some of these objects can be infinite and one would like to 
cancel the infinities between the various terms. In principle all divergent 
objects can be expressed in terms of powers of just a single divergence 
$S_1(\infty)$ times finite terms. This is a rather soft divergence which 
can be regularized rather easily by replacing the infinity temporarily by a 
large integer $N$. In some case one has to worry then about whether objects 
go to infinity like $N$ or like $2N$ in which case one gets additional 
finite contributions as in $S_1(2N) \rightarrow S_1(N)+\ln 2$. These things 
are rather straightforward though.

\appendix
\section{Two loop moments}

The Mellin transform of the coefficient functions $c_2$ from Zijlstra and 
van Neerven~\cite{WillyEllen} 
\begin{eqnarray}
	c_2 & = &
   +\theta(N\minus 3)\ S_{1,-2}(N\minus 3)\ (8/5C_FC_A-16/5C_F^2)
\nn
\IfColor{\textBrown}{}
   +\theta(N\minus 3)\ S_{1,-2}(N\minus 2)\ (-8/5C_FC_A+16/5C_F^2)
\nn
   +\delta(N\minus 2)\ \zeta_3\ (12/5C_FC_A-24/5C_F^2)
\nn
   +\theta(N\minus 2)\ (
   +S_{1}(N\minus 2)\ (8/5C_FC_A-16/5C_F^2)
\nn
   +S_{2}(N\minus 2)\ (8/5C_FC_A-16/5C_F^2)
\nn
\IfColor{\textBlue}{}
   +S_{-4}(N\minus 1)\ (12C_FC_A-24C_F^2)
   +S_{-3,1}(N\minus 1)\ (-8C_FC_A+16C_F^2)
\nn
   +S_{-2}(N\minus 1)\ (8C_FC_A-16C_F^2)
   +S_{-2,-2}(N\minus 1)\ (-24C_FC_A+48C_F^2)
\nn
   +S_{1}(N\minus 1)\ (1585/54C_FC_A-89/27C_Fn_f+5/2C_F^2)
\nn
   +S_{1}(N\minus 1)\ \zeta_3\ (-36C_FC_A+48C_F^2)
\nn
   +S_{1,-3}(N\minus 1)\ (-24C_FC_A+48C_F^2)
\nn
   +S_{1,-2}(N\minus 1)\ (36C_FC_A-72C_F^2)
   +S_{1,-2,1}(N\minus 1)\ (8C_FC_A-16C_F^2)
\nn
   +S_{1,1}(N\minus 1)\ (311/9C_FC_A-26/9C_Fn_f-43C_F^2)
\nn
   +S_{1,1,1}(N\minus 1)\ (22/3C_FC_A-4/3C_Fn_f+8C_F^2)
\nn
   +S_{1,1,-2}(N\minus 1)\ (24C_FC_A-48C_F^2)
   +S_{1,1,1,1}(N\minus 1)\ (24C_F^2)
\nn
   +S_{1,2}(N\minus 1)\ (-22/3C_FC_A+4/3C_Fn_f-4C_F^2)
\nn
   +S_{1,1,2}(N\minus 1)\ (4C_FC_A-32C_F^2)
   +S_{1,2,1}(N\minus 1)\ (-4C_FC_A-24C_F^2)
\nn
   +S_{2}(N\minus 1)\ (-212/5C_FC_A+4C_Fn_f+189/5C_F^2)
\nn
   +S_{1,3}(N\minus 1)\ (12C_FC_A+4C_F^2)
   +S_{2,-2}(N\minus 1)\ (-8C_FC_A+16C_F^2)
\nn
   +S_{2,1}(N\minus 1)\ (-44/3C_FC_A+8/3C_Fn_f+8C_F^2)
\nn
   +S_{2,1,1}(N\minus 1)\ (-24C_F^2)
   +S_{2,2}(N\minus 1)\ (20C_F^2)
\nn
   +S_{3}(N\minus 1)\ (55/3C_FC_A-10/3C_Fn_f-18C_F^2)
\nn
   +S_{3,1}(N\minus 1)\ (8C_FC_A+8C_F^2)
   +S_{4}(N\minus 1)\ (-12C_FC_A+14C_F^2)
\nn
\IfColor{\textRed}{}
   +S_{1}(N)\ (-4639/45C_FC_A+110/9C_Fn_f+337/5C_F^2)
\nn
   +S_{1}(N)\ \zeta_3\ (72C_FC_A-144C_F^2)
   +S_{1,-3}(N)\ (24C_FC_A-48C_F^2)
\nn
   +S_{1,-2}(N)\ (-56C_FC_A+112C_F^2)
   +S_{1,1,-2}(N)\ (-48C_FC_A+96C_F^2)
\nn
   +S_{1,1}(N)\ (-68C_FC_A+4C_Fn_f+84C_F^2)
   +S_{1,1,1}(N)\ (-8C_F^2)
\nn
   +S_{1,3}(N)\ (-24C_FC_A+48C_F^2)
   +S_{2,-2}(N)\ (-16C_FC_A+32C_F^2)
\nn
   +S_{2}(N)\ (74C_FC_A-4C_Fn_f-74C_F^2)
   +S_{2,1}(N)\ (16C_F^2)
\nn
   +S_{3}(N)\ (-20C_FC_A+28C_F^2)
   +S_{3,1}(N)\ (-8C_FC_A+16C_F^2)
\nn
   +S_{4}(N)\ (12C_FC_A-24C_F^2)
\nn
\IfColor{\textBrown}{}
   +S_{1}(N\plus 1)\ (3914/27C_FC_A-488/27C_Fn_f-121C_F^2)
\nn
   +S_{1}(N\plus 1)\ \zeta_3\ (-84C_FC_A+144C_F^2)
\nn
   +S_{1,-3}(N\plus 1)\ (-40C_FC_A+80C_F^2)
\nn
   +S_{1,-2}(N\plus 1)\ (20C_FC_A-40C_F^2)
   +S_{1,-2,1}(N\plus 1)\ (8C_FC_A-16C_F^2)
\nn
   +S_{1,1}(N\plus 1)\ (668/9C_FC_A-68/9C_Fn_f-68C_F^2)
\nn
   +S_{1,1,1}(N\plus 1)\ (22/3C_FC_A-4/3C_Fn_f+36C_F^2)
\nn
   +S_{1,1,-2}(N\plus 1)\ (56C_FC_A-112C_F^2)
   +S_{1,1,1,1}(N\plus 1)\ (24C_F^2)
\nn
   +S_{1,2}(N\plus 1)\ (-22/3C_FC_A+4/3C_Fn_f-32C_F^2)
\nn
   +S_{1,1,2}(N\plus 1)\ (4C_FC_A-32C_F^2)
   +S_{1,2,1}(N\plus 1)\ (-4C_FC_A-24C_F^2)
\nn
   +S_{1,3}(N\plus 1)\ (28C_FC_A-28C_F^2)
\nn
   +S_{2}(N\plus 1)\ (-1909/15C_FC_A+38/3C_Fn_f+646/5C_F^2)
\nn
   +S_{2,1}(N\plus 1)\ (-44/3C_FC_A+8/3C_Fn_f-48C_F^2)
\nn
   +S_{2,1,1}(N\plus 1)\ (-32C_F^2)
   +S_{2,2}(N\plus 1)\ (28C_F^2)
\nn
   +S_{3}(N\plus 1)\ (115/3C_FC_A-10/3C_Fn_f-4C_F^2)
\nn
   +S_{3,1}(N\plus 1)\ (8C_FC_A+24C_F^2)
   +S_{4}(N\plus 1)\ (-12C_FC_A-6C_F^2)
\nn
\IfColor{\textBlue}{}
   +(-S_{1}(N\plus 2)
   +S_{1,-2}(N\plus 2))\ (72/5C_FC_A-144/5C_F^2)
\nn
   +(S_{2}(N\plus 2)
   +S_{3}(N\plus 2))\ (72/5C_FC_A-144/5C_F^2)
\nn
\IfColor{\textRed}{}
   +S_{1,-2}(N\plus 3)\ (-72/5C_FC_A+144/5C_F^2)
\nn
   +S_{3}(N\plus 3)\ (-72/5C_FC_A+144/5C_F^2)
\nn
\IfColor{\textBlack}{}
   -5465/72C_FC_A+457/36C_Fn_f+331/8C_F^2
\nn
   +\zeta_3\ (54C_FC_A-72C_F^2) )  \nonumber
\end{eqnarray}

\end{document}